\documentclass[a4paper,english,aps,prl,twocolumn,floatfix,showpacs,amsfonts,amssymb,superscriptaddress]{revtex4} 
\usepackage{graphics}
\usepackage{amssymb}
\usepackage{amsmath}
\usepackage{overpic}

\newcommand{\be}{\begin{equation}}
\newcommand{\ee}{\end{equation}}
\newcommand{\bea}{\begin{eqnarray}}
\newcommand{\eea}{\end{eqnarray}}

\newcommand{\ybco}{YBa$_{2}$Cu$_{3}$O$_{6+x}$}

\def\d{\delta}

\def\l{\lambda}
\def\th{\theta}

\def\ra{\rightarrow}

\def\pd{\partial}

\def\lb{\label}
\def\pref#1{(\ref{#1})}

\newcount\bozza \bozza=1
\ifnum\bozza=0
\newdimen\shift \shift=-2truecm
\def\lb#1{%
{\label{#1}\rlap{\kern\shift{$\scriptstyle#1$}}}}
\else\def\lb#1{\label{#1}} \fi

\begin{document}

\title{Kosterlitz-Thouless behavior in
layered superconductors:\\ the role of the vortex-core energy}

\author{L.~Benfatto}
\affiliation
{Centro Studi e Ricerche ``Enrico Fermi'', via Panisperna 89/A, I-00184,
  Rome, Italy} 
\affiliation
{CNR-SMC-INFM and Department of Physics, University of Rome ``La
  Sapienza'',\\ Piazzale Aldo Moro 5, I-00185, Rome, Italy}

\author{C.~Castellani}
\affiliation
{CNR-SMC-INFM and Department of Physics, University of Rome ``La
  Sapienza'',\\ Piazzale Aldo Moro 5, I-00185, Rome, Italy}

\author{T.~Giamarchi}
\affiliation{DPMC- MaNEP University of Geneva, 24 Quai Ernest-Ansermet CH-1211
Gen\`eve 4, Switzerland}

\date{\today}

\begin{abstract}
In layered superconductors (SC) with small interlayer Josephson coupling
vortex-antivortex phase fluctuations characteristic of quasi
two-dimensional (2D) Kosterlitz-Thouless behavior are expected to be
observable at some energy scale $T_d$. While in the 2D case $T_d$ is
uniquely identified by the KT temperature $T_{KT}$ where the universal
value of the superfluid density is reached, we show that in a generic
anisotropic 3D system $T_d$ is controlled by the vortex-core energy, and
can be significantly larger than the 2D scale $T_{KT}$. These results are
discussed in relation to recent experiments in cuprates,
which represent a typical experimental realization of layered anisotropic SC.
\end{abstract}


\maketitle

Since the pioneering work of Kosterlitz and Thouless\cite{KT} (KT) on the
so-called KT transition in the two-dimensional (2D) $XY$ model, much
attention has been devoted to the effect of phase fluctuations in quasi 2D
superfluid systems. Thin films are natural candidates for the observation
of KT physics, as the occurence of the ``universal''
(i.e. sample-independent) jump of the superfluid density, measured in
$^4$He superfluid films, or the non-linear $I-V$ characteristic, observed
in thin films of conventional SC.\cite{review} Signatures of KT physics can
be expected also in layered SC with weak inter-layer coupling. A remarkable
example of systems belonging to this class are underdoped samples of
high-$T_c$ SC.\cite{reviewLee} Recently, various experiments ranging from
finite-frequency conductivity,\cite{Corson,Martinoli} Nerst
effect\cite{Wang} and non-linear magnetization\cite{Li} have been
interpreted as signatures of KT phase fluctuations. Nonetheless, any effect
reminiscent of the universal jump of the superfluid density at $T_{KT}$,
which would be the most direct probe of KT physics, failed to be
observed\cite{Lee,Kamal,Panagopoulos,Zuev,Broun}.

Until now, the 2D-3D crossover in anisotropic layered SC has
been discussed mainly within the framework of the anisotropic
3D $XY$ model\cite{Cataudella,Chatto,Friesen,Pierson,Minnaghen}
\be
\lb{xy}
H_{XY}=-\sum_{\langle ij \rangle}J_{ij}\cos (\th_i-\th_j).
\ee
Here $\theta_{i,j}$ is the superconducting phase on two nearest-neighbor
sites $(i,j)$ of a coarse-grained lattice, on the same plane
($J_{ij}=J_{ab}$) or in neighboring planes ($J_{ij}=J_c$).
The energy scales $J_{ab}, J_c$ can be
related to the measured 3D superfluid density $\rho_s$ at
$T=0$ as:
$$
J_{ab}=\frac{\hbar^2d\rho_s^{ab}}{4m}=\frac{\hbar^2 c^2 d}
{\l_{ab}^2 16\pi e^2}, \quad
J_c=\frac{\hbar^2 a\rho_s^c}{4m}=\frac{\hbar^2 c^2a}{\l_c^2 16\pi e^2}
$$
where $\lambda_{ab},\l_c$ represent the in-plane and out-of-plane
penetration depth, respectively, $m$ is the electron mass, $a$ is the
in-plane lattice spacing and $d$ is a transverse length scale (i.e. the
interplane distance) used to define the effective 2D areal superfluid
density $\rho_s^{2d}=d\rho_s$. In a 2D system vortex fluctuations drive
$\rho_s^{2d}(T)$ to zero at $T_{KT}$ given by (we put $\hbar=c=k_B=1$)
\be
\lb{tkt}
\frac{\rho_s^{2d}(T_{KT})}{m}=\frac{8}{\pi}T_{KT},
\ee
where the temperature dependence of $\rho_s^{2d}(T)$ includes also the
effect of other excitations, like long-wavelength phase-fluctuations of the
model \pref{xy} or BCS-like quasiparticles
excitations.\cite{Panagopoulos,reviewLee,noi,noi3} Within the anisotropic
$XY$ model \pref{xy} a finite interlayer coupling $J_c$ cuts off the
logarithmic divergence of the in-plane vortex potential at scales $\sim
a/\sqrt{\eta}$,\cite{Cataudella} where $\eta=J_c/J_{ab}$, so that the
superfluid phase persists above $T_{KT}$, with $T_c$ at most few percent
larger than $T_{KT}$.  \cite{Chatto,Friesen,Pierson} As far as the
superfluid density is concerned, there is some theoretical\cite{Chatto} and
numerical\cite{Minnaghen} evidence that even for moderate anisotropy the
universal jump at $T_{KT}$ is replaced by a rapid downturn of $\rho_s(T)$
at a temperature scale $T_d\simeq T_{KT}$.

However, recent measurements of $\rho_s(T)$ in strongly underdoped {\ybco}
(YBCO) samples\cite{Zuev,Broun} (with large $\eta \sim 10^{-4}$
anisotropy\cite{Broun,Hosseini}), showed that no downturn of $\rho_s(T)$ is
observed at the KT temperature defined by Eq.\ \pref{tkt}, but eventually
at a scale $T_d\approx T_c$.\cite{Zuev} Analogously, recent measurements of
the phase-fluctuations diamagnetism in underdoped Bi2212\cite{Li} revealed
that the phase correlation length $\xi$ above $T_c$ can be fitted with the
typical KT law, provided that the effective KT temperature is few kelvin
smaller than $T_c$. In both cases, by looking at the system from below or
above $T_c$, it appears that the typical temperature scale where vortex
fluctuations become relevant is always near $T_c$, regardless of the value
\pref{tkt} of the $T_{KT}$ of the pure 2D case.

In this paper we analyze the role played by the interlayer coupling and the
vortex-core energy at the crossover from 2D KT to 3D superconducting
behavior in layered SC. In particular, we focus on the
behavior of the superfluid density below $T_c$ and of the
correlation length above $T_c$. We carry out a
renormalization group (RG) analysis using the mapping between the thermal
metal-SC KT transition in 2D and the quantum
metal-insulator transition in the 1D sine-Gordon model.\cite{Tbook}
Indeed, a similar model has been studied in Ref.\
[\onlinecite{thierry}] to investigate the superfluid-insulator transition
in optical lattices of 1D boson chains, where the tunneling inter-chains
amplitude plays the same role of the Josephson coupling in layered
SC.
%
We show that in the presence of a finite interlayer coupling the superfluid
density looses its universal character. The jump in $\rho_s(T)$ at $T_{KT}$
observed in the 2D case is replaced by a downturn curvature at a
temperature $T_d$ which depends on the vortex-core energy
$\mu$. While in $XY$ models, where $\mu$ is fixed by the in-plane coupling
$J_{ab}$ (see Eq.\ \pref{eqmu}), $T_d\simeq T_{KT}$, in the general
case the ratio $T_d/T_{KT}$ increases as $\mu/J_{ab}$
increases. Analogously, by approaching the transition from above, the
increasing of the phase-fluctuation correlation length is controlled by the
scale $T_d$ instead of the $T_{KT}$ of the pure 2D system.  Based on these
results, we argue that the various experimental data in cuprates concerning
KT behavior can be reconciled if $\mu$ is larger than the typical
$XY$ value.

Let us first recall briefly the basic features of the KT transition using
the analogy with the quantum 1D sine-Gordon model,\cite{review,Tbook}
 defined as:
\be
\lb{sg}
H_{sg}=\frac{v_s}{2\pi}\int_0^L dx \left[K(\pd_x\theta)^2+\frac{1}{K}
(\pd_x\phi)^2 + \frac{2g_u}{a^2}\cos(2\phi)\right].
\ee
Following standard definitions,\cite{Tbook} $\theta$ and $\pd_x\phi$
represent two canonically conjugated variables for a 1D chain of length
$L$, with $[\theta(x'),\pd_x \phi(x)]=i\pi\d(x'-x)$, $K$ is the
Luttinger-liquid (LL) parameter, $v_s$ the velocity of 1D fermions, and
$g_u$ is the strength of the sine-Gordon potential.  For $g_u=0$ the
action of the model \pref{sg} can be simplified by integrating 
$\phi$ and rescaling $\tau\rightarrow v_s\tau$, so that
\be
\lb{s0}
S_0=\frac{K}{2\pi}\int dx d\tau \left[(\pd_x\theta)^2+(\pd_\tau\theta)^2
\right],
\ee
equivalent to the gradient expansion of the model \pref{xy}, with
$\tau$ as the second spatial dimension.  Note that in the Hamiltonian
notation \pref{sg} the coefficient of the dual field is $1/K$, 
while the rotational in-plane symmetry of the model \pref{xy}
is recovered in the action \pref{s0}. Besides the long-wavelength phase
fluctuations present in Eq.\ \pref{s0}, vortex configurations are possible,
which require $\oint \nabla \theta =\pm 2\pi$ over a closed loop.  Since
$\phi$ is the dual field of the phase $\theta$, a $2\pi$ kink in the field
$\theta$ is generated by the operator $e^{i2\phi}$,\cite{Tbook} i.e. by the
sine-Gordon potential in the Hamiltonian \pref{sg}. The correspondence
between the quantum 1D and the classical 2D system is then completed by using:
\be
\lb{defk} K\equiv \frac{\pi J_{ab}}{T}, \quad g_u\equiv y=2\pi e^{-\beta
\mu}, \ee
where $y$ is the vortex fugacity. 
In the 2D $XY$ model \pref{xy} (with $J_c=0$) one has a single energy
scale given by $J_{ab}$, so that the 
vortex-core energy $\mu$ is given by:\cite{KT,review}
\be
\lb{eqmu}
\mu_{XY}=\pi J_{ab} \ln(2\sqrt{2}e^\gamma)\simeq 1.6 \pi J_{ab},
\ee
where $\gamma$ is the Euler's constant. However, $\mu$ depends in general
on the details of the microscopic superconducting model under
consideration, so it will be taken as a {\em free} parameter in the
following, while the value \pref{eqmu} will be used just for the sake of
comparison with the $XY$ model \pref{xy}. It is worth noting that the
limitations of the $XY$ model as an effective phase-only model have been
pointed out in Ref. [\onlinecite{noi}], as far as the role of the
phase-interaction terms beyond Gaussian level are concerned.
The effect of the interlayer coupling $J_c$ of Eq.\ \pref{xy} can
be  incorporated in the sine-Gordon model \pref{sg} as an interchain
hopping term, so that the full Hamiltonian becomes:
\be
\lb{ham}
H=\sum_m H_{sg}[\phi_m,\theta_m]
-\frac{v_sg_{J_c}}{\pi a^2}\sum_{m\atop{m'=m\pm 1}}
\int_0^L dx \cos[\theta_m-\theta_{m'}],
\ee
where $m$ is the chain (layer) index and $g_{J_c}\equiv\pi
J_c/{T}$.  We derived the perturbative RG equations for the couplings
of the model \pref{ham} by means of the operator product expansion, in
close analogy with the analysis of Ref.\ [\onlinecite{thierry}].  Under RG
flow an additional coupling $g_\perp$ between the phase in neighboring
chains is generated:
\be
\lb{gf}
\frac{g_\perp}{2\pi}\sum_{\langle m,m'\rangle} \int dx \left[
-K(\pd_x\theta_m)(\pd_x\theta_{m'})+\frac{1}{K}
(\pd_x\phi_m)(\pd_x\phi_{m'})\right].
\ee
The superfluid coupling $K_s$ is defined, as usual,\cite{Minnaghen} as the
second-order derivative of the free energy with respect an infinitesimal
twist $\delta$ of the phase, $\pd_x \theta_m\rightarrow
\pd_x\theta_m-\delta$. The interlayer term \pref{gf} contributes as $K\ra
K(1-ng_\perp)$ to the current-current coefficient, where $n=2$ is
the number of nearest-neighbors chains (layers). Thus, the in-plane
stiffness $J_s$ is defined as:
\be
\lb{defks}
K_s=K-nKg_\perp,  \quad J_s=\frac{\rho_s^{2d}}{4m}=\frac{K_s T}{\pi}
\ee
The full set of RG equations for the couplings $K,K_s,g_u,g_{J_c}$
reads:
\bea
\lb{eqk}
\frac{dK}{d\ell}&=&2g_{J_c}^2-K^2g_u^2,\\
\lb{eqgu}
\frac{dg_u}{d\ell}&=&(2-K)g_u,\\
\lb{eqks}
\frac{dK_s}{d\ell}&=&-g_u^2K_s^2,\\
\lb{eqgj}
\frac{dg_{J_c}}{d\ell}&=&\left(2-\frac{1}{4K}-\frac{K_s}{4K^2}\right)g_{J_c},
\eea
with $\ell=log(a/a_0)$, where $a_0,a$ are the original and RG rescaled
lattice spacing, respectively.  Observe that for $g_{J_c}=0$ the first two
equations reduce to the standard ones of the KT transition,\cite{KT} with a
fixed point at $K=2, g_u=0$, and $K_s$ coincides with $K$.
Thus, one sees that at $K>2$ the $g_u$ coupling is irrelevant, the quantum
1D system is a LL and the vortex-antivortex pairs are bound in the
classical 2D system. At $K<2$ the $g_u$ term is relevant, the $\phi$ field
is locked in a minimum of the $\cos(2\phi)$ potential and the 1D system is
an insulator. In the classical case this corresponds proliferation of
vortexes (large vortex fugacity) in  the metallic
phase. The physical superfluid stiffness is given\cite{review} by the
asymptotic value of the running coupling $K_s(\ell)=K(\ell)$. Thus, $J_s$
is finite below $T_{KT}$, since $K(\ell)$ flows to a finite value (in
particular $K(\infty)=2$ at $T_{KT}$, in accordance to Eq.\ \pref{tkt}),
and it goes to zero above the transition, since $K(\ell)$ scales to
zero. The KT temperature is defined by the highest temperature where
$K(\infty)=2$, and it is given (at small $g_u$) by 
$K(T)-2=2g_u(T)$, which yields $T_{KT}\simeq \pi J_{ab}/2$.
\begin{figure}[htb]
\includegraphics[scale=0.3,angle=-90]{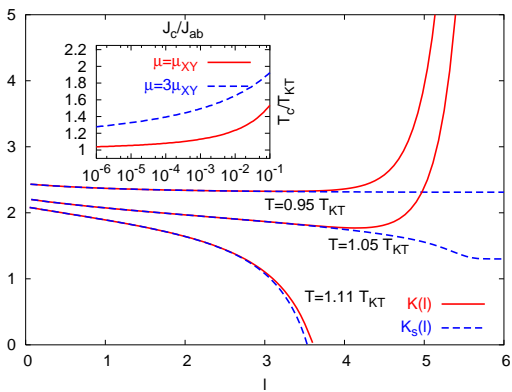}
\caption{(Color online) RG flow of the couplings $K(\ell)$ and $K_s(\ell)$
  at various temperatures for $\mu=\mu_{XY}$ and $\eta=10^{-4}$. Inset:
  critical temperature $T_c$ as a function of $\eta$ for a
  bare stiffness $J_{0}(T)=J_{ab}(1-T/4J_{ab})$ (see text).}
\label{Fig-1}
\end{figure}

As an initial value $g_{J_c}\neq 0$ is considered, the interchain
(interlayer) coupling increases under RG,\cite{Chatto,Friesen,Pierson}
leading to larger values of the LL parameter $K(\ell)$ and stabilizing the
metallic 1D phase.  However, when the initial $g_u$ coupling is
sufficiently large the second term in the r.h.s of Eq.\ \pref{eqk}
dominates and $K(\ell)$ goes to zero, leading to the insulating 1D phase.
In the classical 2D analogous the effects of $g_{J_c}$ are easily readable
trough the behavior of $K_s$, which is controlled by the $g_u$ coupling
alone. Whenever $K(\ell)$ scales to large values the $g_u$ coupling is
irrelevant and $K_s$ flows to a constant, see Fig.\ 1. This effect
guarantees the persistence of the superfluid phase in a range of
temperature above $T_{KT}$. Indeed, the initial decrease of $K_s(\ell)$ is
cut off at a finite length scale by the interlayer coupling, which brings
again $K(\ell)$ to large values and $g_u(\ell)$ to zero, giving a finite
asymptotic value of $K_s(\ell)$.  As the temperature increases further and
the $g_u$ term dominates, both $K$ and $K_s$ scale to zero, the scaling
dimension of $g_{J_c}$ becomes negative and one observes a ``layers
decoupling'' above $T_c$.\cite{Friesen,Pierson}
\begin{figure}[htb]
\includegraphics[scale=0.3,angle=-90]{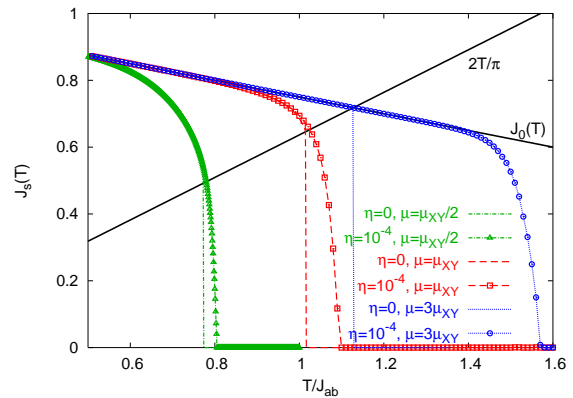}
\caption{(Color online) Temperature dependence of
 $J_s(T)$ in 2D (lines) and in the layered 3D case (symbols). Here
 $J_0(T)=J_{ab}(1-T/4J_{ab})$. The $T_{KT}$ is identified by
 the intersection between $J_s(T)$ and the straight line $2T/\pi$. The
 results for $\mu\leq\mu_{XY}$ show a rapid downturn of $J_s(T)$ at
 $T_d\simeq T_{KT}$. As $\mu$ increases $T_d$ increases as well, so that
 at $T_{KT}$ no effect is observed in $J_s(T)$ reminiscent of the jump
 present in 2D.}
\label{Fig-2}
\end{figure}

The critical temperature $T_c$ is defined by the vanishing of the
$K_s(\ell\ra\infty)$. Alternatively, to account for the perturbative
character of the RG equations, one can compute $J_s$ by stopping the RG
flow at the scale $\ell^*$ where $g_{J_c}$ is of order one.\cite{nota} The
two definitions are equivalent, and lead to the estimate of the critical
temperature $T_c$ reported in the inset of Fig.\ 1.  For the sake of
completeness, we also added a temperature dependence of the bare couplings,
using $J_{0}(T)=J_{ab}(1-T/4J_{ab})$, as due to long-wavelength phase
fluctuations in the $XY$ model \pref{xy},\cite{Minnaghen,noi} and we keep
the ratios $\eta=J_c/J_{ab}$ and $\mu/J_{ab}$ fixed.  For $\mu=\mu_{XY}$
the calculated values of $T_c/J_{ab}$ show a remarkable quantitative
agreement with Monte Carlo simulations on the anisotropic $XY$
model.\cite{Minnaghen} The fact that larger values of $\mu$ lead to a
larger critical temperature has a direct counterpart on the temperature
dependence of the superfluid stiffness $J_s(T)$, as we show in Fig.\ 2.  As
one can see, when $J_c=0$ we recover the standard jump
of $J_s$ at $T_{KT}$, which is easily identified as the temperature where
the curve $J_s(T)$ intersects the line $2T/\pi$, according to Eqs.\
\pref{tkt} and \pref{defks} above. As $\mu$ increases the $g_u$ coupling
decreases, and the renormalization of $J_s(T)$ with respect to $J_{0}(T)$
below $T_{KT}$ becomes negligible. As soon as a finite interlayer coupling
is switched on, the jump of $J_s(T)$ at $T_{KT}$ disappears and it is
replaced by a rapid bending of $J_s(T)$ at some temperature $T_d$. However,
while for $\mu\leq\mu_{XY}$ $T_d$ coincides essentially with $T_{KT}$, for
a larger vortex-core energy $T_d$ rapidly increases and approaches the
temperature $T_c$ estimated above.
\begin{figure}[htb]
\includegraphics[scale=0.3,angle=-90]{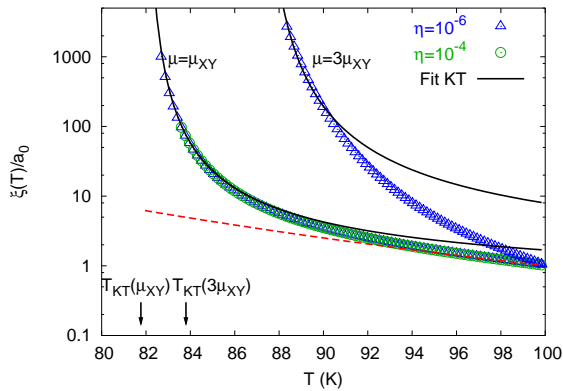}
\caption{(Color online) Correlation length above $T_c$ for different values
  of $\eta$ and $\mu$. Each curve stops at $T_c$, which increases as $\eta$
  increases. We take $J_{0}(T)=J_{ab}(1-T/T_{MF})$, with $J_{ab}=300$ K and
  $T_{MF}=100$ K, as appropriate for Bi221.\cite{Lee} The dashed line is
  $\xi\sim e^{\beta\mu(T)/2}$, expected to hold far above $T_c$. The solid
  lines are fit with the KT functional form (see text), with $T_{d}=T_{KT},
  c=0.25, b=0.9$ for $\mu=\mu_{XY}$ and $T_{d}=86.8$ K, $c=0.4, b=1.21$ for
  $\mu=3\mu_{XY}$.}
\label{Fig-5}
\end{figure}

These results offer a possible interpretation of the experiments in
underdoped YBCO.\cite{Broun}], where $\eta\sim 10^{-4}$\cite{Hosseini},
but the measured
$J_s(T)$ goes smoothly across the $T_{KT}$ estimated from Eq.\
\pref{tkt}. We calculated $J_s(T)$ as done in Fig.\ 2 (taking into account
also the measured linear depletion at low $T$). 
Using a large vortex-core energy, i.e. $\mu=6\mu_{XY}$, we found
that $J_s(T)$ shows no signature of a rapid downturn at
$T_{KT}$, and goes to zero near to the measured $T_c$.  Observe that we did
not consider the effects of disorder, which can
also smear out the KT transition, as measurements in thin
films\cite{Martinoli,armitage} could suggest.

A similar separation between $T_{KT}$ and $T_d$ is observed in the behavior
of the correlation length above $T_c$.  Since the quantity experimentally
accessible is the vortex density $n_V$, given at the RG scale
$\ell=\ln(a/a_0)$ by $n_V(\ell)=e^{-\beta\mu(\ell)}/a^2=g_u(\ell)/(2\pi
a^2)$, we define the correlation length $\xi$ as $\xi^{-2}\equiv
n_V(\ell_s)$, where $a_s=a_0e^{\ell_s}$ is the length scale where
$K_s(\ell_s)$ vanishes above $T_c$.  The behavior of $\xi(T)$ for different
values of $\mu$ and $\eta$ is reported in Fig.\ 3, using parameter values
appropriate for Bi2212 compounds.  Far above $T_c$ $a_s\approx a_0$, so
that $\xi$ scales as $a_0/\sqrt{g_u(T)/2\pi}$, as shown by the dashed
line. As $T$ approaches $T_c$ $\ell_s$ increases and $\xi$ shows the
exponential increase reminiscent of the KT behavior in 2D.\cite{KT}
However, while in 2D $\xi$ diverges at $T_{KT}$, a finite $J_c$ cuts off at
$T_c$ the increasing of $\xi$, since below $T_c$ $g_u$ becomes irrelevant
and $K_s$ flows to a finite value. Nonetheless, the behavior of $\xi$ above
$T_c$ is still reminiscent of the KT behavior, $\xi_{KT}\sim a_0c\,
\exp(b/\sqrt{T/T_{d}-1})$, with $c,b$ of order one, provided that $T_{KT}$
is replaced by a proper scale $T_{d}$ slightly smaller than $T_c$. Once
again, while for $\mu=\mu_{XY}$ $T_{d}\sim T_{KT}$, as the vortex-core
energy increases $T_{d}$ becomes significantly larger than $T_{KT}$ and
approaches $T_c$. This behavior is consistent with recent 
experiments in Bi2212 compounds.\cite{Li}

In summary, we analyzed the phase-fluctuations contribution to the 2D KT-3D
crossover in strongly anisotropic layered SC. Using a RG approach, we
showed that a finite interlayer coupling can shift the temperature scale
$T_d$ of vortex unbinding away from the KT temperature $T_{KT}$ of the pure
2D case.  Indeed, $T_d$ is essentially controlled by the vortex-core
energy, and it coincides with $T_{KT}$ only when $\mu\lesssim \mu_{XY}$, as
within the standard $XY$-model \pref{xy}. When applied to cuprates, our
findings suggest that in these systems $\mu$ is definitively larger than
expected in the $XY$ model, even though still of order of the in-plane
stiffness. The consequences are twofold. First, the lack of any signature
of KT behavior in $J_s(T)$ at $T_{KT}$ does not rule out the possibility
that phase-fluctuation effects play a role in these systems.  Second,
$\mu>\mu_{XY}$ is not inconsistent with microscopic theories which
associated $\mu$ to the energy scale of the superfluid stiffness instead of
that of the superconducting gap, which would be far too large compared to
$J_s$ in underdoped samples.\cite{reviewLee} 

\acknowledgments

We ackonowledge useful discussion with S. Caprara, M. Cazalilla and
M. Grilli. This work was supported in part by the Swiss NSF under MaNEP
 and division II.


\begin{thebibliography}{20}
\bibitem{nota} In analogy with Ref.\ [\onlinecite{thierry}], we used
$g(\ell^*)=0.5$.
\bibitem{note-anis} For simplicity we neglected the (much
smaller) in-plane $a-b$ anisotropy, which is expected to have minor
effects.
\end{thebibliography}
\end{document}